\documentclass[lettersize,journal]{IEEEtran}
\usepackage{amsmath,amsfonts}

\usepackage{array}
\usepackage[caption=false,font=normalsize,labelfont=sf,textfont=sf]{subfig}
\usepackage{textcomp}
\usepackage{stfloats}
\usepackage{url}
\usepackage{verbatim}
\usepackage{graphicx}
\usepackage{subfig}
\usepackage{cite}
\usepackage{algorithm}
\usepackage{algorithmic}
\usepackage{multirow}
\usepackage{xcolor}
    \newcolumntype{L}{>{\raggedright\arraybackslash}X}
\usepackage{tabularx}
    \newcolumntype{L}{>{\raggedright\arraybackslash}X}
\newcommand\blfootnote[1]{%
  \begingroup
  \renewcommand\thefootnote{}\footnote{#1}%
  \addtocounter{footnote}{-1}%
  \endgroup
}
\begin{document}

\title{Deep Q-Learning Assisted Bandwidth Reservation for Multi-Operator Time-Sensitive Vehicular Networking

}

\author{Abdullah Al-Khatib, Albert Gergus, Muneeb Ul Hassan, Abdelmajid Khelil, Klaus Mossner, Holger Timinger
}

\maketitle

\begin{abstract}
Very few available individual bandwidth reservation schemes provide efficient and cost-effective bandwidth reservation that is required for safety-critical and time-sensitive vehicular networked applications. These schemes allow vehicles to make reservation requests for the required resources. Accordingly, a Mobile Network Operator (MNO) can allocate and guarantee bandwidth resources based on these requests. However, due to uncertainty in future reservation time and bandwidth costs, the design of an optimized reservation strategy is challenging. In this article, we propose a novel multi-objective bandwidth reservation update approach with an optimal strategy based on Double Deep Q-Network (DDQN). The key design objectives are to minimize the reservation cost with multiple MNOs and to ensure reliable resource provisioning in uncertain situations by solving scenarios such as underbooked and overbooked reservations along the driving path. The enhancements and advantages of our proposed strategy have been demonstrated through extensive experimental results when compared to other methods like greedy update or other deep reinforcement learning approaches. Our strategy demonstrates a 40\% reduction in bandwidth costs across all investigated scenarios and simultaneously resolves uncertain situations in a cost-effective manner. 
\end{abstract}


\section{Introduction} 
\IEEEPARstart{I}{n} recent years, there has been a significant increase in the number of intelligent vehicles equipped with powerful onboard processing capabilities \cite{Nvidia}. Researchers are exploring ways to leverage these onboard computing resources for applications such as self-driving \cite{chen2021energy},\cite{cheng2021multiagent}. \blfootnote {Abdullah Al-Khatib, Abdelmajid Khelil and Holger Timinger are with the Institute for Data and Process Science, Landshut University of Applied Sciences, Germany (e-mail: {Abdullah.Al-Khatib, Abdelmajid.Khelil, Holger.Timinger}@haw-landshut.de).

Klaus Mossner is with the Professorship for Communications Engineering, Technical University Chemnitz, Germany (e-mail: Klaus.Moessner@etit.tu-chemnitz.de).

Muneeb Ul Hassan is with the School of Information Technology, Deakin University, Australia (e-mail: muneebmh1@gmail.com).

Albert Gergus is with the Department of Control Engineering and Information Technology, Budapest University of Technology and Economics, Hungary (e-mail: g.albert95@hotmail.com).}These applications are typically computation-intensive, safety-critical, and time-sensitive, requiring immediate action and reaction to ensure safety. However, the limited onboard computing resources of a single vehicle may not be sufficient to handle the demands of these applications. To overcome this, application data can be offloaded to centralized cloud servers or edge cloud servers via 5G Vehicle-to-Infrastructure (V2I) connections \cite{chiang2016fog}. Such applications demand ultra-low-latency and ultra-reliable communication/network resources (bandwidth) with deterministic guaranteed access to surrounding edge computational resources \cite{al2024resources}. The pattern followed by vehicular safety applications is typically bursty demand, characterized by high bandwidth requirements for a shorter time period to access edge services in hazardous situations \cite{meneguette2021vehicular}.

To ensure the required Quality of Service (QoS) for vehicular safety applications, there has been a growing suggestion for bandwidth reservation which guarantees timely access to network resources. Centralized reservation, where the Mobile Network Operator (MNO) sets aside bandwidth to meet the future needs of various QoS classes is a common approach in the mobile network field. This approach is commonly referred to as network-side reservation \cite{al2020priority, al2021bandwidth}. However, this approach does not provide a deterministic but a probabilistic guarantee to access the network resources for individual vehicles. This may lead to failures in Safety-critical Vehicular Applications (SVA), where it is important to ensure that vehicles have access to the necessary bandwidth to perform critical tasks. Vehicle-side reservation has shown to be a more efficient solution, where vehicles reserve the necessary resources in advance, and make reservations based on their specific requirements and resource cost, rather than relying on the MNO to allocate resources on their behalf
\cite{niyato2008competitive, chen2020edge, zang2019filling,al2022optimal, zang2021soar, al2024optimizing, al2024blockchain}. Another focus of the vehicle-side approach is the economical aspect, which is being minimized by the expenditure for guaranteed access of the network upon the reservation. 

From the viewpoint of business, MNOs have various traditional pricing options for allocating bandwidth. These involve fees for network service reservation (i.e. subscription) and charges for real-time requests. Additionally, MNOs may offer different purchasing packages for users to choose from based on their consumption needs, such as the Pay-As-You-Go (PAYG) option \cite{AmazonW} or a long-term upfront fee for reserved bandwidth (i.e. monthly subscription) \cite{Atandt}. However, these pricing strategies may not always effectively manage peak-time and real-time network conditions while ensuring sufficient QoS for SVA.

Recently, dynamic pricing has emerged as a promising solution for resource management in edge computing \cite{han2018dynamic, liao2021intelligent, luong2016data}. This method adjusts prices in real-time based on network conditions, making it suitable for managing peak-time congestion and ensuring QoS. However, the majority of vehicle-side reservation methods only factor in fixed prices or PAYG models and overlook the dynamic pricing set by MNOs \cite{niyato2008competitive, chen2020edge, zang2019filling, zang2021soar}. This neglect of dynamic pricing may result in overpriced reservations or insufficient resources. An instance of utilizing this method can be observed in Amazon Web Services (AWS), which has introduced a feature called spot pricing \cite{amazonspot}. This approach is not exclusive to AWS, as several companies, including MTN, China Telecom, and Uninor, adopt time-dependent pricing for bandwidth resources. This feature dynamically adjusts prices, often on an hourly or minute-by-minute basis, in response to changes in supply and demand \cite{lin2020backup, sen2013survey}. In a prior study \cite{al2022optimal}, we showed that vehicles can optimize bandwidth costs and mitigate the potential risks associated with dynamic pricing and resource unavailability from MNOs through carefully placed reservation requests that we refer to as smart request.

Despite the benefits of smart reservation requests, still there are challenges in dealing with uncertainties present in real-world situations, such as unpredictable mobility that may affect reservation times and locations. The reservation time can either be “exactbooking” if it remains unchanged or subject to uncertainties such as underbooking or overbooking. Underbooking scenario occurs when the reserved time for each 5G-NR gNodeB (gNB) Base Station falls short of demand, while overbooking leads to the underutilization of resources. To minimize costs, vehicles must continuously update their reservations based on different scenarios discussed in \cite{al2022heuristic}. While this method can be expanded to use for multiple operators, it yields suboptimal results as shown in the results. Furthermore, relying solely on the update of the reservation time at the moment of entering the coverage area of each gNB is insufficient. Such a strategy disregards changes in time during the gNB, which can lead to disruptions in the reservation process. To ensure efficient resource utilization, reservation updates must take place when there is a change in both time and cost. Otherwise, unnecessary updates can lead to paying cancellation fees, which may increase the overall cost. This paper explores efficient frequent reservation updates to ensure reliable resource availability in uncertain situations by regularly checking the reservation time and cost, collecting offers from multiple MNOs, and performing re-reservations to minimize cost while avoiding unused resources.

Problems with cost minimization of resource reservation are typically non-convex nonlinear and NP-hard \cite{chaisiri2011optimization}. Dynamic pricing introduced a new challenge: Vehicles should find the optimal strategy to find the best time to update the reservation in order to minimize the bandwidth cost.
Also, vehicles may enter and leave the coverage of gNB Base Stations due to their mobility, which introduces new challenges and uncertainty in determining the actual reservation time. This can be inefficient in providing resources in highly dynamic environments \cite{olariu2019survey},\cite{meneguette2018avarac} and may compromise guaranteed and reliable resource provisioning \cite{ghazizadeh2015reasoning}.
To address this, vehicles need to solve underbooking and overbooking scenarios as well at the lowest possible bandwidth reservation cost.

Artificial intelligence has gained attention in recent years as a solution to complex problems, such as resource management in wireless networks \cite{zhang2023dynamic, chen2018optimized, he2017deep}. The Bandwidth update reservation presents challenges due to the lack of prior knowledge for optimal algorithm, timing, and actions. However, using supervised learning-based models is complicated due to its need for labeled data. The problem can be represented in the form of a Markov Decision Process (MDP) for which the Deep Reinforcement Learning (DRL) is well-suitable to solve it. DRL can learn the relationship between the state and action and is effective in complex and interactive environments. In contrast to traditional greedy-based methods, RL-based methods, such as DRL, aim to maximize long-term benefits. This makes DRL suitable for complex problems, such as reservation updates, where the effects of a decision may not be immediately apparent but are beneficial in the long run. The most widely used RL method is the Deep Q-Network (DQN) algorithm, but it suffers from Q-value overestimation. To address this shortcoming, Double Deep Q-Network (DDQN) has been proposed \cite{van2016deep} by decoupling action selection and Q-value estimation. We employed Double Deep Q-Learning (DDQL), a DRL that can learn the optimal strategy directly for both challenges: Updating reservations for cost minimization, and solving the underbooking and overbooking in a cost-effective way. To address these challenges, the implementation should utilize a time-discretized learning environment, wherein the interval is segmented into multiple, discrete timeslots. During each timeslot, the available price data is provided to the model, allowing it to determine the optimal decision. The major contributions of our article can be summarized as follows:
\begin{itemize}
\item We formulate the problem of multi-operator switching in vehicular systems as an MDP problem. We take into account the temporal changes of the MNO price, the MNO cancellation cost, and the variable costs of reservation in different problem scenarios, particularly underbooking and overbooking over the reservation time.
\item We adopt a DRL agent that utilizes the DDQL algorithm to find the optimal strategy for both challenges simultaneously: Updating the reservations to minimize bandwidth costs and resolving underbooking and overbooking scenarios in a cost-effective way.
\item For a stable learning process of the DRL agent, we develop an effective MDP environment with states, action space, and reward function that is suitable for the considered multi-operator and different reservation update scenarios.
\item Using simulations based on the historial dataset of Amazon spot prices~\cite{amazonspot, al2022heuristic}, we show that each vehicle that is supported by a DLR bandwidth reservation agent in the edge environment will attain a large reduction in bandwidth cost.
\end{itemize}

The remaining sections of the article have been arranged as follows: Section II reviews relevant work on cost-effective resource reservation. Section III covers the system model and corresponding scenarios. Section IV provides a formulation for the bandwidth reservation problem as MDP. In Section V, we propose bandwidth reservation update strategies and the implementation of the DQL model in this work. In Section VI, we carry out a comparison of the performance of the proposed model with state-of-the-art methods. Section VII concludes the article and discusses future work.

\section{Related Work}
Many studies have explored resource reservation problems using early protocols and models like Resource Reservation Protocol (RSVP), and Internet Integrated Service model \cite{yu2012resource}.  
However, most research has focused on network-side resource reservation in mobile networks \cite{al2020priority, al2021bandwidth}. Few studies have considered the economic implications of vehicle-side reservation, such as resource reservation with different pricing strategies. This is becoming a growing area of interest in edge computing \cite{niyato2008competitive,chen2020edge,zang2021soar,zang2019filling,al2022optimal, al2022heuristic, al2024blockchain}. 

\begin{table*}[ht!]
    \centering
    \caption{The summary of approaches to the cost-effective resource (communication/computation) reservation request/update problems.}
    \begin{tabularx}{\linewidth}{ |p{1.5cm}|p{0.5cm}|p{1.5cm}|p{2.2cm}|L|L| }
    \hline
       \textbf{Request type} & \textbf{Ref.} & \textbf{Approaches} & \textbf{Mechanism Type} & \textbf{Functioning of Mechanism} & \textbf{Objectives}
     \\ \hline\hline
         Reservation request (provisioning) &\cite{cao2016share} & Game theory & Auction & Competition-based onsite negotiations resources request through bid models &
    -Creating competition between users to discover price and leverage the demand and supply resources 
    \\ \cline{2-2} \cline{3-6}
    & \cite{chen2020stackelberg} & Game theory & Stackelberg game & Cooperative game model between MNO and the users & -Achieving the maximum revenue of MNO under user’s budget constraints
    \\ \cline{2-6}
    & \cite{zang2019filling}& Immediate request & Individual-scenario & Deterministic and randomized online reservation & -Reducing the cost of both plane reservation and PAYG
    \\ \cline{2-2} \cline{3-6}
    & \cite{zang2021soar}& Immediate request & Broker-scenario & Smart online aggregated reservation (SOAR) & -Minimizing the cost of multiple users after aggregating their tasks
    \\ \cline{2-6}
    & Our prior work \cite{al2022optimal}& Advanced request & Smart advanced reservation request & Cost-effective advanced smart reservation & -Reducing the risk of dynamic price and failure to access the resource in advance \newline -Minimizing reservation costs through the selection of the optimal time for a request 
   \\ \hline
   Reservation update / \mbox{re-reservation} & \cite{chaisiri2011optimization} & Immediate update & Conventional onsite  reservation update & Optimizing the costs of resource provisioning in reservation phase and update phase & -Minimizing the reservation and on-demand provisioning cost under uncertainty (demand and price)
   \\ \cline{2-6}
   & Our prior work \cite{al2022heuristic} & Single-operator advanced update & Reservation update strategy & Heuristic greedy continuous updated bandwidth reservation strategy & -Minimizing the amount of time and the costs of the resources are reserved with different scenario e.g. exactbooking, underbooking, and overbooking
   \\ \cline{2-6}
   & \cite{zhang2023dynamic} & Single-operator reservation update & Single-objective reservation update & DeepReserve dynamic reservation of edge servers based on observed demands and make reservation decisions to maximize system utility & -Maximize the system utility and optimize resource utilization \newline
   -Increase the probability of successful requests
   \\ \cline{2-6}
    & This work & Multiple-operator reservation update & Multi-objective reservation update strategy &
 Efficient strategy for updating the reservation at the optimal time & -Ensuring reliable resource provisioning in uncertain environments while minimizing costs by updating at the optimal time \newline -Minimizing the total cost, which includes both reservation and cancellation costs, across different scenarios by timely updating for more cost-effective options
   \\ \hline
    \end{tabularx}
\end{table*}

Resource reservation can generally be divided into two phases: i) reservation request, where users request a specific amount of bandwidth for a specific period and location from the MNO, ii) reservation update, where users modify an existing reservation, such as increasing the amount, changing the duration, or adjusting the priority. 
Generally, most studies related to reservation request mainly put emphasis on the former mentioned onsite competition \cite{cao2016share},\cite{bajaj2015spectrum} or immediate request mode. The main difference between those two types of requests is that in competition requests,  users compete for the resource through various game theoretic ways, such as auctions, Stackelberg game, etc. \cite{cao2016share, chen2020stackelberg, bajaj2015spectrum}. This results in only a limited number of winners acquiring the resources, leading to a risk of failure for some users and violation of QoS. Furthermore, onsite requests frequently exhibit volatile pricing and inherent inequity due to the stochastic nature of resource availability and demand, compounded by the rigorous standards for SVA. In contrast, immediate requests, as discussed in  \cite{niyato2008competitive}, focus on addressing the challenge sharing the available spectrum between multiple secondary users and a primary user. In this scenario, primary users offer pricing information to secondary users, allowing them to reserve spectrum and optimize their utility. Chen et al. \cite{chen2020edge} developed an approach based on meta-learning to assist in reserving resources for computing with the goal of minimizing the cost of using edge services. Zang et al. worked over proposing a smart online reservation framework to minimize the cost of reserving resources for an individual user \cite {zang2019filling} or multiple users \cite{zang2021soar}. Based upon their settings, the approaches discussed above on reservations mainly operate on an immediate request basis. As a result of limited resources, the corresponding vehicles needs to carry out the reservation of schedule in much advance in order to ensure they are able to acquire the necessary resources on time. As a consequence, the immediate requests entail high cost vales alongside low guarantees. Planning reservations efficiently is a challenge as users lack knowledge about cost trends and available resources, making it difficult to ensure cost-effectiveness.

Motivated by challenges incurred by competition and immediate request, a smart request solution  \cite{al2022optimal} has been introduced. This solution enables the advanced reservation of mobility locations at specific time intervals, achieving commendable cost-effectiveness and time efficiency. In reality, the scheduled reservation requires to be updated accordingly in order to consider the uncertain vehicle mobility nature, as well as the uncertainty surrounding the reservation of future times and prices. This uncertainty may lead to the over- or underutilization of restricted resources. For instance, a study by \cite{chaisiri2011optimization} has investigated the issues associated with on-site updates, which focus on minimizing the initial reservation and on-demand provisioning cost under uncertainty in demand and price.  Other methods only rely on updating the reservation at the time of entering the coverage area \cite{al2022heuristic}, lacking the ability to compare earlier prices for updating in a timely manner. Another method presented by Zhang et al is DeepReserve \cite{zhang2023dynamic}, which dynamically reserves edge servers for connected vehicles. The key objective is to optimize the allocation of edge computing resources to enhance the performance of connected vehicle applications. This paper explores how this dynamic reservation system can contribute to improving the overall performance and responsiveness of connected vehicle services. However, these algorithms are satisfactory for single-operator scenarios but significantly underperform when generalized for multiple-operator environments, where continuously comparing the available MNO prices and finding the optimal time for multi-objective updating is of higher importance. In our recent work \cite{al2025bandwidth}, we addressed the challenge of cost-effective reservation request in a multi-operator environment using Dueling Deep Q-Learning. By introducing a novel area-wise approach and leveraging synthetic data generated with the Temporal Fusion Transformer (TFT), we demonstrated up to 40\% cost savings over baseline scenarios.
The contemporary approaches to resource reservation request and update problems are listed in Table I.

In summary, we emphasize the critical need for a cost-effective and timely multi-operator reservation update approach, which includes an update strategy as a fundamental component in a dynamic pricing vehicular network environment. While previous literature has not extensively covered this aspect, our work makes a novel contribution in this direction. Our study aims to provide reliable resource provisioning in uncertain situations while minimizing costs across various scenarios. We believe that our work can serve as a valuable stepping stone for researchers in this field, addressing the gaps and advancing the state-of-the-art in reservation update strategy.
\begin{figure}[t]
    \centering
    \includegraphics [width=0.45\textwidth]{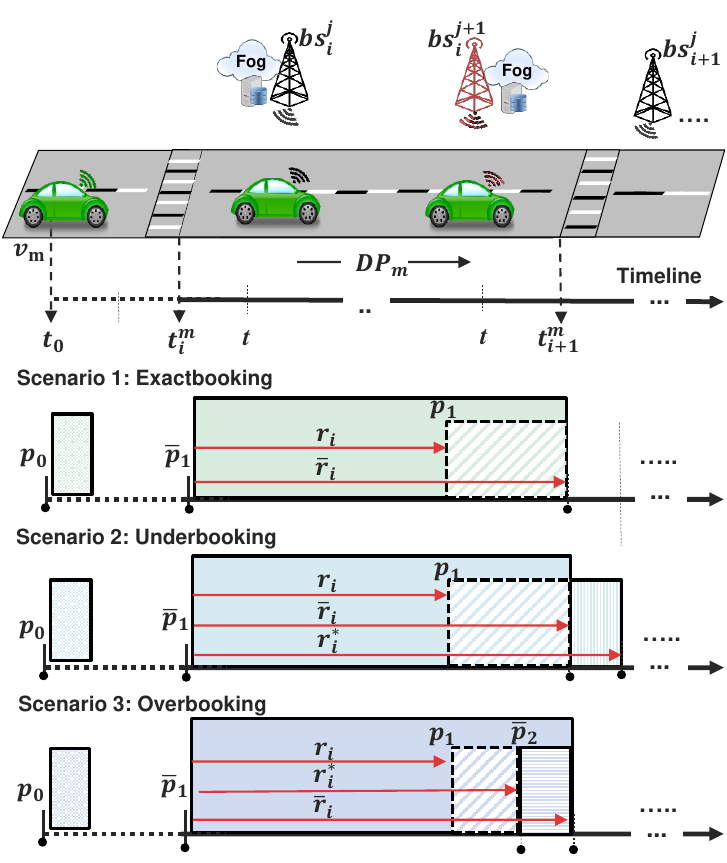}
    \caption{System model and an example of exactbooking, underbooking and overbooking scenarios.}
    \label{fig:systemmodel_last}
\end{figure}
\section{SYSTEM MODEL AND PRELIMINARIES}

We consider that a vehicular network is composed of two main components: vehicles and gNB Base Stations (BSs), provided in Fig. \ref{fig:systemmodel_last}. BSs can take the form of either Road Side Units (RSUs) or Macro Base Stations (MBSs). According to their network range the driving path (DP)\footnote[1] {Our assumption is that the path, meaning the road segment sequence which leads to the destination, is predetermined and known in advance  (e.g., the scheme in \cite{nadembega2012destination}.} can be split into $N$ road segments\footnote[2] {We refer to road segments as a road stretch in between a handoff point and intersection or in between two intersections. We designate each segment by a location pair  $(a, b)$. For this, we refer the point at which a road intersects the boundary of a BS coverage area is termed a handoff point.}, each having $M$ number of operating MNOs, which facilitate a wireless connection between the edge routers and the affiliated vehicles of the core network. We assume that each BS covers one segment\footnote[1] {E.g., the scheme in \cite{nadembega2014mobility}.} and within each segment, all MNOs have a dedicated BS, accounting for a set of $bs_i^j \; i \in 1...N , j \in 1...M$ BSs at multiple locations along the DP. For simplification, we refer to the set of all available BSs at road segment $i$ as $BS_i$. To facilitate vehicular applications with stringent latency demands, a Fog/Edge server node is embedded within the wireless network infrastructure of each BS. The important notations are listed in Table II.

\subsection{Reservation Request/Update Model}
MNOs offer resources through subscriptions \cite{Atandt}, where the users have to pay a variable price in order to reserve the specific bandwidth for the specific interval of time, irrespective of either they are utilizing the specified resource or not. Contrarily, if allocated time is insufficient, the vehicle may request additional time (in the form of a partial reservation). Our scenario involves vehicles operating on a time slot-based system, with the assumption that every specific vehicle can have a connection with a single MNO network in a given time slot $t$. Furthermore, the agent has access to all MNOs' presently available prices. Two main stages have been considered: Reservation request stage and the stage of reservation update. In the former, the vehicle plans the initial reservations, and in the latter, an RL agent is used to update reservations where needed. 

\textit{1. Initial Reservation Request:} During this stage, vehicle $v_m$ at time $t_0$ makes the initial reservation request to the initially lowest-cost MNOs for each road segment along the DP. This request specifies the bandwidth time period $\Delta t_m$ required for $v_m$ to finish its intended $DP_m$. The time period $\Delta t_m$ is further divided into  $\overline{r_i}=[t_{i}^m, t_{i+1}^m] \; i \in 1...N$ reservation duration intervals, the time the vehicle $v_m$ passes through the coverage area\footnote[1] {The diameter of the coverage areas of a BS is approximately 900m, similar to \cite{soh2006predictive}, and it is typically visible in downtown areas of metropolises.} of a road segment covered by $BS_i$.
These $\overline{r_i}$ reservation durations depend on the $DP_m$ length covered by the corresponding set of $BS_i \; i \in 1...N$ and the vehicle speed. We assume that the future $DP_m$ and route information like BS coverage area $\overline{r_i}$ are computed and reported based on the information from navigation system. Accordingly, we only consider the time-based bandwidth reservation. Moreover, we assume that the same amount of requested bandwidth (in unit of Mb/s) is needed at different BS locations. This is possible because the vehicle can determine its resource requirements by taking into account the values of data, applications, and the task models, which further aid in predicting the necessary bandwidth using various methods pretty similar to those given in the work \cite{chen2020edge}.

\textit{2. Reservation Update\footnote[2] {In our model, we make the same assumption similar to various reservation schemes already available, which is that every associated BS has a specified and fixed capacity of bandwidth \cite{soh2006predictive}, \cite{choi2002adaptive}. In the scenario, if a vehicle is underbooked and cannot obtain bandwidth due to all resources being booked, the MNO may resort to alternative methods, such as overbooking, which is discussed in detail in \cite{liwang2022overbooking}.}:} This stage involves monitoring the available resource reservation costs with the RL agent in order to make reservation changes. The vehicle $v_m$ starts its utilization of the initially reserved resources as per his arriving times at a handoff point at $t_{i}^m$. The ending point of the current set of $BS_i$ (which is the handoff time to the subsequent $BS_{i+1}$) is denoted by $t_{i+1}^m$. Before this point, vehicle signs whether the previously reserved $\overline{r_i}$ is sufficient or should be changed. Also, the vehicle monitors the currently available prices of the available MNOs, searching for a lower-cost operator to connect to\footnote[3] {The prices are accessible, such as on AT\&T \cite{Atandt} and Amazon \cite{amazonspot}.}. To handle these, we assume that the currently utilized MNO provides a partial cancellation strategy\footnote[4] {We follow the cancellation strategy employed by Microsoft Azure\cite{Microsoft}.}. 

\begin{table}[t]
    \caption{List of Important Notations.}
    \begin{tabularx}{\linewidth}{ |p{1.5cm}|L| }
    \hline
       \textbf{Notations} & \textbf{Definition}
       \\ \hline\hline
       $v_m$ & Vehicle using the bandwidth reservation strategy
       \\ \hline
       $DP_m$ & The planned driving path of vehicle $v_m$
       \\ \hline
       $BS_i$ & All BSs of the $i^{th}$ road segment
       \\ \hline
       $bs_i^j$ & BSs of the $i^{th}$ road segment for the $MNO_j$
       \\ \hline
       $t_0, T$ & Initial and final timestep for $DP_m$
       \\  \hline
       $\Delta t_m$ & Required time for the vehicle to go through the $DP_m$
       \\ \hline
       $t_{i}^m$, $t_{i+1}^m$ & The reservation start and end at the currently used MNO
       \\ \hline
       $\overline{r}_i$ & Originally planned reservation duration period between $t_{i}^m$ and $t_{i+1}^m$
       \\ \hline
       $r_i$ & Actual reservation duration after partial cancelation
       \\ \hline
       $r_i^{*}$ & Reservation duration after updating for solving underbooking or overbooking
       \\ \hline
       ${p}_i$ & Earlier reserved bandwidth price rate that is currently paid
       \\ \hline
       $\overline{p}_i$ & Available bandwidth prices for updating at the MNOs of the current $BS_i$
       \\ \hline
       $\overline{p}_{i+1}$ & Available bandwidth prices for updating at the MNOs of the next $BS_{i+1}$
       \\ \hline
       $c_i^p$ & Partial cancelation fee
       \\ \hline
       $c^p$ & Per minute cancelation fee
       \\ \hline
       $\overline{c}_i^p$ & Under/Overbooking cancelation fee
       \\ \hline
       $c^{\textit{reservation update}}$ & The individual cost of the different objectives: updating \\
       $c^{\textit{under}}, c^{\textit{over}}$ & the reservation for bandwidth fee minimalisation \\ 
       & and solving the under and overbooking
       \\ \hline
       $C^{\textit{exact}} C^{\textit{under}}$, $C^{\textit{over}}$ & Total cost of exactbooking, underbooking and overbooking scenarios
       \\ \hline
       $b$ & Variable to show the booking scenario (-1 for underbooking, 0 for exactbooking, 1 for overbooking)
       \\ \hline
       $\mathcal{S}, \mathcal{A}, \mathcal{R}, \mathcal{P}_r$ & State-, action-, reward- and state transition space of the Markov decision process
       \\ \hline
       $s_t, a_t, R_t$ & State, action and reward at timestep $t$
       \\ \hline
       $R(s_t, a_t)$ & Reward function
       \\ \hline
       $p_{max}, p_{min}$ & Possible maximum and minimum prices
       \\ \hline
       $P(s_t, a_t)$ & State transition function
       \\ \hline
       $G_t$ & Cumulative reward collected during an episode
       \\ \hline
       $\mathcal{Q}$ & Q function of the action Q-network
       \\ \hline
       $\hat{\mathcal{Q}}$ & Q function of the target Q-network
       \\ \hline
       $\tau$ & Soft update weight for the target agent's weights
       \\ \hline
    \end{tabularx}
\end{table}

If the vehicle is offered a bandwidth price for the remaining part of the current BS, then it can decide to cancel the reservation at the previous higher price and make a new reservation at the new lower price. In case of overbooking, the vehicle reaches the handoff point earlier than planned, hence, the entire remaining unnecessary bandwidth at a particular road segment covered by $BS_i$ is canceled and a reservation is made at $BS_{i+1}$ in its place. Alternatively, in case of underbooking, the vehicle would not reach the next road segment covered by $BS_{i+1}$ by the previously planned $t_{i+1}^m$ so the vehicle should cancel the first part reserved at $BS_{i+1}$ and reserve additional bandwidth at the current $BS_i$ (Fig. \ref{fig:systemmodel_last}). Additionally, we modeled the possibility of a DP consisting of the random mix of exact-, under-, and overbooking (called mixedbooking). The vehicle can decide whether to update or wait at each time slot of the reservation update stage. 

\subsection{Markov Decision Process Model}
An MDP can be categorized as a mathematical model which is used to carry out representation of decision-making in the specific scenarios where the results are influenced by both chance and the actions of a decision-maker. It comprises of a states set, reward function, actions, and the rule for determining the next state, i.e. the state transition function, $(\mathcal{S}, \mathcal{A}, \mathcal{R}, \mathcal{P}_r)$. The goal in an MDP is mainly to figure out a policy, which is basically the set of rules in order to decide which action needs to be taken in every state, so that one maximizes the expected sum of rewards over time. MDPs are useful to carry out various decision-making tasks in a widespread number of applications. We propose to use an MDP to model and solve the bandwidth reservation problem for vehicular systems. We use a DQL algorithm for policy decision-maker so that we can minimize cost of multi-operator bandwidth reservation problem.

The goal of the agent is to decide when and how to update the reservation upon underbooking and overbooking problems (and their mix) and minimize costs on the basis of the current stage.

\section{Formulation of a Bandwidth Reservation Problem as Markov Decision Process}
\label{MDP}
In order to apply DQL algorithm for the bandwidth reservation problem, we need to formulate it as an MDP \cite{sutton2018reinforcement}, which consists of a decision-maker (the RL agent) and an environment. In this application, the vehicle ($v_m$) makes decisions on the bandwidth reservation, so we denote it as the RL agent in our framework. The environment in the DP usually consists of a set of road segments with $N$ BS, $BS_i, i=1,...,N$, each having $M$ MNOs and BSs, that the vehicle moves through.

MDP framework is defined by a 4-tuple \((S,A,R, P_r)\). At every step of time \(t\), RL agent carries out interaction with environment, observes the state value \(s_t \in S\), and performs selection of an action \(a_t \in A\) which further determines the reward \(R_t \sim R(s_t,a_t)\) for the selected action in the given state, and the next state \(s_{t+1} \sim P(s_t, a_t)\). As described in \cite{zhang2020deep}, both the state and action space are finite. The ultimate objective of RL agent is to maximize the cumulative reward, which is the total sum value of all the received rewards. 

If the RL agent-environment interaction can be naturally broken into sequences, the task is called episodic. Then, the cumulative reward, denoted as $G_t$, is the sum of all rewards received during an episode, starting from time-step $t+1$ up to the final time-step $T$. Therefore, $G_t = R_{t+1}+R_{t+2}+...+R_T$.

In the following four subsections, we carry out formulation of action space, transition function, state space, and reward signal for bandwidth reservation. The reward signal formulation is applicable to all scenarios of exactbooking, underbooking, and overbooking and their mix. 

\subsection{Problem Formulation}
Considering that the initial reservations can be suboptimal, to guarantee the necessary resources at the lowest possible price, the reservation will need to be updated as needed. A update is required when the initial reserved bandwidth period is insufficient or when there are available bandwidth prices that save costs. In case of an update the previously reserved bandwidth should be canceled and a new one should be made during which a cancellation fee stipulated by the MNO, which can be calculated by: 
$c_{i}^{p}=c^{p}(\overline{r_i}-r_i)$, where $c^{p}$ is the per minute cancelation fee and $(\overline{r_i}-r_i)$ is the bandwidth time canceled. The bandwidth reservation update problem can be split into two subproblems. 

First, the vehicle can check for lower available unit prices provided by different MNOs for the currently used BS. If a lower price is available the vehicle can cancel the remaining part of the previous reservation, make a new reservation, and continue the rest of its course area covered by the BS with the new price until the handoff point or the next cancelation. The main aim is to reduce the total reservation cost for every road segment $BS_i$ across the entire $DP_m$. The condition for updating the reservation to be advantegeous is: 

\begin{equation}
    \overline{p_i}(\overline{r_i}-r_i) + c_{i}^{p} < p_i(\overline{r_i}-r_i)
\label{eq:state}
\end{equation} 

Meaning that we should pay less at the new price for the remaining part of the BS plus the cancelation fee than we would pay at the current price. At every timestep our agent can decide whether to update to a new price or keep our reservation at the current price.

Second, the vehicle must update its reservation in the case of underbooking or overbooking. In order to determine whether an update is needed in the future, the agent is complemented by a navigation system. This system can calculate and update the anticipated time of the next handoff point, and determine the amount of over- or underbooking. In case of underbooking, the vehicle will arrive to the handoff point later than it has initially booked, so the vehicle needs to cancel the underbooked period of the next $BS_{i+1}$ and re-reserve additional bandwidth for the current $BS_i$. However, in the case of overbooking, the vehicle will arrive to the handoff point earlier than initially reserved, so it needs to cancel the overbooked period at the current $BS_i$ and re-reserve for the next $BS_{i+1}$ and all subsequent BSs up to $BS_N$. In both cases, the reservation update has to be made until reaching the handoff point. Our goal is to update at the lowest cost. We propose an RL agent to address this problem: this agent will decide at each timestamp whether to update the reservation or wait for a lower price. 

\subsection{State Space}
The RL agent performance depends on the choice of the space state. The state accommodates the information about the environment available to the RL agent, and represents the basis for decision making \cite{sutton2018reinforcement}. In the bandwidth reservation service, at time step $t$, state should contain the currently reserved bandwidth price $p_i$, the available prices $\overline{p}_i$ for current $BS_i$, and the available prices for the next $BS_{i+1}$, provided by the available MNOs. The vehicle will signal in advance to the agent if an update is needed to solve underbooking or overbooking as $b$ (-1 for underbooking, 0 for exactbooking, 1 for overbooking). The same system will calculate and send the agent the $t_i'$ timesteps left of the current BS - the time until reaching the next handoff point. Therefore, the state at timestep $t$ is given by:

\begin{equation}
    s_t = [p_i, \overline{p}_i, \overline{p}_{i+1}, b, t_i']_t, \quad i = 0,...,N
\label{eq:state}
\end{equation}

\subsection{Action Space}
Actions are taken by the RL agent, vehicle $v_m$, in order to maximize the reward -- incur the minimum total reservation cost. By the definition of DQL algorithm, we define the space of all possible actions. The algorithm should learn to select the best action in the specific state on the basis Q-value of that specific state-action pair. 

At the time-step $t_0$, vehicle $v_m$ receives initial bandwidth reservations $\overline{r}_i = [t_i^m, t_{i+1}^m], \quad \forall BS_i, \quad i \in{1,...,N}.$. The vehicle will calculate whether the current bandwidth at $BS_i$ should be updated based on the information from the navigation system, and inform the agent how many timesteps are left for updating. For the first task the goal of the agent is whether it is profitable to change to re-reserve at one of the available MNOs at $BS_i$. To address the second task, the agent can update to solve the underbooking overbooking at one of the currently available prices, or wait for a better price.


\begin{equation}
\begin{aligned}
    A_t = 
    \begin{cases}
        & \text{do\_nothing}, \\
        & \text{solve\_underbooking}, \\
        & \text{solve\_overbooking}, \\
        & \text{change\_to\_lowest\_price\_mno},\\
        & \quad i = 0,...,N
    \end{cases}
    \label{eq:action}
\end{aligned}
\end{equation}

\subsection{Reward Signal}
Our objective is to reduce the total cost of reservation in the three scenarios as a vehicle $v_m$ moves through $DP_m$. These three scenarios are exactbooking, underbooking, and overbooking. The reward signal in RL reflects the cost incurred in each scenario, and the reward should be the larger the smaller are the costs. These three scenarios, and therefore three components of the reward function, are mutually exclusive. To enable smooth gradient flow in DQL, the reward signal should be reshaped in the function with maximum and minimum. In order to standardize the prices, we utilized the maximum and minimum price values and then reshape the reward function. 

In the following subsection, we explain how to construct the reward function for the exactbooking scenario. The same procedure is applicable for the underbooking and exactbooking scenarios. Each BS can be exactbooking, underbooking, or overbooking scenario, and a DP can consist of homogeneously just one type or their random mix: make cost functions for only a single BS and at the R(t) sum them together.

\subsubsection{Exactbooking Scenario}
In the exactbooking scenario, the reserved bandwidth $\overline{r}_i$ corresponds to the demand. However, if we find a lower unit price $p_i$, we will cancel the remaining part of the reservation $\overline{r}_i$ made at the old price $\overline{p}_i$, and make a new reservation $r_i$. We minimize the cost function of the whole $DP_m$, consisting of $N$ number of BSs:

\begin{equation}
    C^{\textit{exact}} = \sum_{i=1}^N \overline{p}_i r_i + p_i (\overline{r}_i-r_i)+c^p_i
\label{eq:exactsum}
\end{equation}

The action policy should satisfy a number of conditions $\overline{C}_i$. As we work over determining the optimal action corresponding to each state, we assign a large negative reward to every action that violates any of conditions $\overline{C}_i$ that are dependent on action choice. Such action is trying to update for underbooking or overbooking, while it the current BS is exact booked. In this case we add a large negative part to the reward. 

\subsubsection{Underbooking Scenario}
If the initial bandwidth reservation period is smaller than needed, we have underbooked computation time. For underbooking scenario, the reward function is similar to the reward function strategy used for the exactbooking scenario, but it is complemented with the price of solving the underbooking. Based on the formula of the cost function for underbooking scenario:

\begin{equation}
    C^{\textit{under}} = \sum^N_{i=1}{\overline{p}_i\left(r^*_i-{\overline{r}}_i\right)+\ {\overline{p}}_ir_i+p_i\left({\overline{r}}_i-r_i)+c^p_i+{\overline{c}}^p_i\right.}
\label{eq:undersum}
\end{equation}

Where $({\overline{r}}_i-r^*_i)$ is the amount of underbooking: the difference of the real handoff time and the previously planned handoff time, and $\overline{c}^p_i=c^{p}(r^*_i-{\overline{r}}_i)$ is the cost of canceling the initial reservation for the underbooked time period at the next $BS_{i+1}$. As in the case of exactbooking scenario, the actions that violate defined constraints have to be prevented, namely missing to update to solve underbooking. In case of constraint violation, a large negative penalty is added to the existing reward in order to significantly decrease the Q-value, and prevent missing this action.

\subsubsection{Overbooking Scenario}
If the vehicle would arrive earlier to the next handoff point than the bandwidth time we have reserved, an overbooking scenario occurs. Similarly to underbooking, the cost of overbooking consists of the cost of exactbooking and solving the overbooking scenario:

\begin{equation}
    C^{\textit{over}} = \sum^N_{i=1}{\overline{p}_{i+1}\left({\overline{r}}_{i}-r^*_{i}\right)+\ {\overline{p}}_ir_i+p_i\left({\overline{r}}_i-r_i)+c^p_i+{\overline{c}}^p_i\right.}
\label{eq:oversum}
\end{equation}

The overbooked period is $({\overline{r}}_i-r^*_i)$: the difference of the previously planned handoff time and the real handoff time. The cost to cancel the overbooked period is $\overline{c}^p_i=c^{p}(({\overline{r}}_i-r^*_i)$, the cancelation cost of the initial reservation for the overbooked time period at the current $BS_{i}$. In the overbooking scenario we also add a large negative penalty if the agent misses updating to solve overbooking.

\subsubsection{Mixed Booking Scenarios}
As in the mixed booking scenario either exactbooking, underbooking or overbooking can occur at each BS, the reward depends on the base stations' scenario: either the cost of exactbooking, underbooking, or overbooking. In this case, either of the constraints of the exactbooking, underbooking, and overbooking- holds depending on the BS scenario, and has to be prevented by a penalty. 

\begin{figure}[t]
    \centering
    \includegraphics[width=9cm, height=8cm]{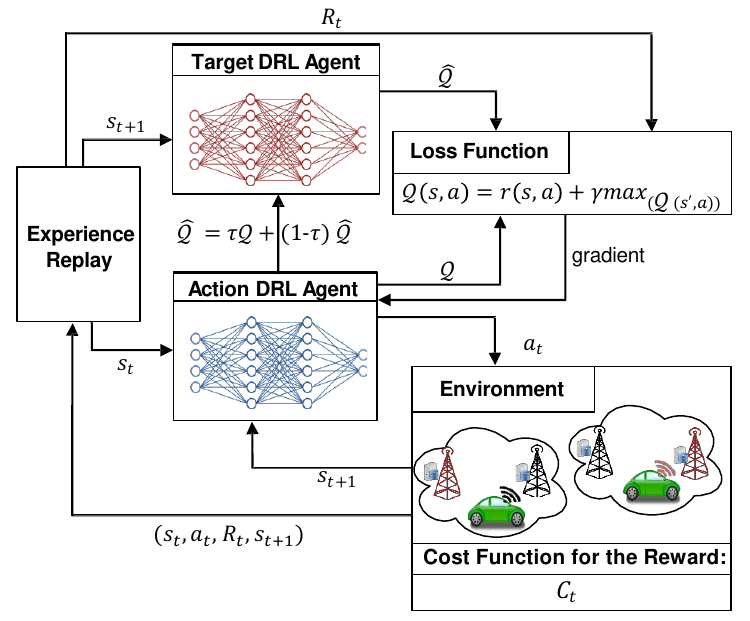}
    \caption{An illustration of the proposed double Q-learning algorithm for bandwidth reservation update strategy.}
    \label{fig:reward}
\end{figure}

\section{Deep Q-Learning for Bandwidth Reservation Update Strategy} 

RL can be used to design a bandwidth reservation update strategy for networked vehicular applications by training an agent to learn a policy for making bandwidth reservations that maximizes a reward signal. The agent can learn a policy for making reservation updates by interacting with the MDP and receiving rewards based on the performance of the reservation strategy, as illustrated in Fig. \ref{fig:reward}.

\subsection{Deep Q-Learning}
In modern networking systems, efficient bandwidth reservation is critical for optimal network performance. To address this challenge, DQL can be applied for dynamically updating network resources and improving performance by learning from experience. While approximating Q-function via deep neural networks, DQL can handle high-dimensional state and action spaces and can adapt to changing network conditions, making it a promising approach for addressing bandwidth reservation in modern networks.

The agent receives a reward for successfully updating the necessary bandwidth at a low price, minimizing the cost of the bandwidth reservation problem, and a penalty for reserving at a high price. The agent would also receive a reward for solving the underbooking and overbooking problem at lower price, and a penalty for not solving it, or solving it at a higher price.

Q-Learning is a type of reinforcement learning algorithm to train an agent how to select actions within a specific environment in order to achieve maximum reward \cite{watkins1989learning}. The primary element of Q-Learning is the Q-value function, to estimate the expected return, the inverse of the expected future payment for the specific action for staying in the current MNO, changing to a new one or solving the under/overbooking in a given state. By using the Bellman equation to update the Q-values on the basis of the reward received and subsequent state, the agent step by step improves its decision-making process.

\begin{equation}
    \mathcal{Q}(s, a) = r(s, a) + \gamma max_(\mathcal{Q}(s', a))
\label{eq:q-value}
\end{equation}

Where $r(s, a)$ is the normalized inverse the  cost according to the current BSs state: $C^{exact}$ for exact booked and $C^{over}$ or $C^{under}$ for over and underbooked BSs respectively. To enable agents' ability to explore new states and for balancing exploration and exploitation we use epsilon-greedy algorithm. This involves choosing some actions randomly with probability epsilon to experiment out-of-policy actions to minimize the bandwidth cost.

\begin{equation}
a = \begin{cases}
\text{random action} & \text{with probability } \epsilon 
\\ \arg\max_{a'} Q(s, a') & \text{with probability } 1 - \epsilon
\end{cases}
\label{eq:egreedy}
\end{equation}

We also employ experience replay, which involves recording the previous environment interactions that an agent has encountered, to increase the sampling efficiency and stability of training. 

 A DQL approximates (state, action) Q-value function with the help of a deep neural model. This technique is useful when dealing with a large state space, as is the case with multiple mobile MNOs that need to be represented. With a state as input, the outputs number in the given neural network corresponds to the number of possible actions, and the value of each output represents the Q-value for the (state, action) pair. Afterwards, an action with maximum Q-value for the specific state is chosen.
 
This algorithm tends to be unstable because the determined policy can not only fail to converge, but actually diverge, that is the reason we use this algorithm. The double refers to the presence of the target neural network. Here, to calculate the Q-value to learn, instead of using the trained \textit{action agent's} action with the best $\mathcal{Q}$ value, we use a \textit{target agent's} $\mathcal{\hat{Q}}$ value's best action. This helps in reducing the Q-values overestimation. In order to avoid the divergence of the two models, we regularly update the \textit{target agent} with the weights of the \textit{action agent}.

\begin{equation}
    \mathcal{\hat{Q}} = \tau\mathcal{Q} + (1-\tau)\mathcal{\hat{Q}}
\label{eq:softupdate}
\end{equation}

The pseudocode of the DDQL can be found in \cite{van2016deep}. 
\begin{algorithm}[t]
    \begin{algorithmic}[1]
    \STATE Initialize  Replay memory
    \STATE Initialize  Q-network $\mathcal{Q}$ and the target Q-network $\hat{\mathcal{Q}}$ with arbitrary weight $\theta$ and $\hat{\theta}$
    \STATE Initialize $Q: \mathcal{X} \times \mathcal{A} \rightarrow \mathbb{R}$ arbitrarily
    \FOR{Episode: j=0 to n}
        \STATE Initialize t = 0
        \FOR{BS: $i$=0 to $N$}
            \FOR{\textit{timestep} in $r_i$}
                \STATE Choose action with epsilon-greedy \eqref{eq:egreedy} \\
                \STATE Calculate the cost function with \text{\eqref{eq:exactsum}, \eqref{eq:undersum} or \eqref{eq:oversum}}
                \STATE Scale $C_t$ to receive $R_t$
                \STATE Collect reward $R_t$ and observe state $s_{t+1}$
                \STATE Store the transition $(s_{t}, a_{t}, R_{t}, s_{t+1})$ 
                \STATE t = t + 1
                \STATE \textbf{Until} $s_{t+1}$ is not terminal state
            \ENDFOR
        \ENDFOR
        \STATE Sample random experience minibatch of transitions $(s_{t}, a_{t}, R_{t}, s_{t+1})$
        \STATE Calculate the target Q value: \\ $y_j=R_j+\gamma \mathcal{Q} (s_{t+1}, argmax_a(\mathcal{\hat{Q}}(s_{t+1}, a_{t+1};\hat{\theta})); \theta)$
        \STATE Perform a gradient descend step on $(y_j - \mathcal{Q}(s_t, a_{t}; \theta))^2$ with regard to $\theta$
        \STATE At every $C$ step update $\mathcal{\hat{Q}} = \tau\mathcal{Q} + (1-\tau)\mathcal{\hat{Q}}$
    \ENDFOR
    \end{algorithmic}
\caption{Training DDQL Bandwidth Reservation Update Strategy}
\label{alg:q-learning}
\end{algorithm}

\subsection{DDQL for the Bandwidth Reservation Update Problem} 
DQL is a suitable algorithmic choice because of present uncertainties and the time-efficiency requirement. The unpredictable mobility changes of the vehicle can affect future reservation times. Moreover, uncertainty arises due to the nature of bandwidth prices. The bandwidth reservation update has to be time-efficient to guarantee resource provision for mentioned vehicular application. In comparison to classical methods, running a neural network agent requires a higher level of computational resources. However, we implemented a model that utilizes a smaller network architecture to reduce the computational demands, enabling the agent to operate effectively on low-end embedded systems.

In Section \ref{MDP}, state space, action space, reward signal, and state transition function are defined. These definitions are necessary for the application of DQL and for implementing the bandwidth reservation algorithm \eqref{alg:q-learning}. After initialization in Lines 1-3, we use the action agent to generate episodes. These agents will go through predefined routes with $N$ BSs, that can be further split to timesteps. For all timesteps during the episode we use epsilon-greedy algorithm (function \eqref{eq:egreedy}) to select the actions in Line 8 from the possible actions:

\begin{equation}
a_t = \begin{cases} 
            \text{switch to MNO with lower price}
            \\ \text{reservation update for under/overbooking}
            \\ \text{do nothing}
        \end{cases}
\end{equation}

Then the algorithm calculates the cost for the current timestep according to the current booking scenario (Line 9). To gather the rewards we invert the cost, scale it to standard distribution for better convergence, and store it along with the transition to the experience replay memory in Lines 9-12. 

After every episode we execute a training step. During this process we sample a minibatch, consisting of a number of transitions from the replay memory (Line 16), and make a gradient descent step (Line 18) to the Q-value calculated at Line 17, using the target agent's $\mathcal{\hat{Q}}$ action for the action agent's Q-value function. Finally, we periodically make a soft update step on the target agent with the action agent's weights according to \eqref{eq:softupdate}. 

Our DDQL method employs a fixed number of $n$ episodes, each consisting of at most 400 timesteps (\ref{exp_setting_paragraph}). Utilizing early-stopping schema, in the worst-case scenario our training exhibits linear time complexity, with a computational complexity of $\mathcal{O}(n)$ iterations.
For the implementation concerning bandwidth reservation, we adapt two optimization goals in the pseudocode of the Q-learning algorithm.

\subsubsection{Reservation Update for Cost Minimalization}
The first optimization goal is to minimize the bandwidth reservation cost by switching to one of the available MNOs. The agent has access to the available MNOs' currently provided prices and chose between staying at the previously reserved price or updating to the new price in one of the available MNOs. The goal of the agent is to minimize the cost of the bandwidth reservation problem. If it does not update the reservation, then the vehicle will continue to pay at the previously reserved price. Upon update, there is a fee for partially canceling the previous reservation, and the vehicle should also reserve a new bandwidth, of which price will be the cost until reaching the handoff point of the current road segment $BS_i$ at $t_{i+1}^m$ or until the next cancellation. Furthermore, as there is a s fee for every update, frequent updates would yield a high overall cost. To minimize the total cost the agent should find and update only at the best prices. In the specific case of exact booking, our sole objective is to minimize costs with reservation updates, and thus, the cost associated with the exact booking is equivalent to the cost we aim to minimize within this objective (\ref{eq:exactsum}).

\begin{equation}
    c^{\textit{reservation update}} = C^{\textit{exact}} = \sum_{i=1}^N \overline{p}_i r_i + p_i (\overline{r}_i-r_i)+c^p_i
\label{eq:reservation_update}
\end{equation}

\subsubsection{Reservation Update for Underbooking or Overbooking}
In the second problem, based on the navigation system the vehicle determines whether exactbooking, underbooking, or overbooking are happening in order to choose the corresponding reward function. On top of the former problem (\ref{eq:reservation_update}), the goal here is to solve the non-exactbooking problem with the lowest possible cost. This can be done based on currently available prices for either the current $BS_i$ (for underbooking) or the next $BS_{i+1}$ (for overbooking). The $t'_i$ flag indicates the timesteps left to solve the underbooking or overbooking problem. The agent can choose at each timestep, whether to update and reserve additional bandwidth at one of the available MNOs or to wait for lower prices at the next time step. 
Therefore, in these scenarios, the costs associated with addressing underbooking and overbooking (\ref{eq:under}, \ref{eq:over}) are added to the cost of the initial exact booking (\ref{eq:reservation_update}). This combination yields the total cost for these scenarios (\ref{eq:undersum}, \ref{eq:oversum}).

\begin{equation}
    c^{\textit{under}} = \sum^N_{i=1}{\overline{p}_i\left(r^*_i-{\overline{r}}_i\right)+ \overline{c}}^p_i.
\label{eq:under}
\end{equation}

\begin{equation}
    c^{\textit{over}} = \sum^N_{i=1}{\overline{p}_{i+1}\left({\overline{r}}_{i}-r^*_{i}\right)+\overline{c}}^p_i
\label{eq:over}
\end{equation}

\section{PERFORMANCE EVALUATION}

In this section, we present the results of detailed experimental studies. To demonstrate the efficiency of proposed method, we compare our algorithm with various existing bandwidth reservation schemes in vehicular settings.

\subsection{Experimental Settings} \label{exp_setting_paragraph}
To reduce the cost of a reservation, the agent continuously searches for the lowest priced unit at a time-slot, initiating from the time of initial reservation request ($t_0$) up to the destination. The cost at each timestep (BS) is calculated as the sum of the bandwidth prices and the cancellation fees, which are then accumulated for each BS in the episode to carry out algorithms performance evaluation. The length of a BS is randomly varied between 10 and 20 minutes with a 30-second timestep resolution, and the duration of an episode is determined by a random number of BSs ranging from 3 to 10. 

Our training used several key hyperparameters to optimize the learning process. We used Adam optimizer with a learning rate set to $10^-3$, with a weight decay of $10^-6$ to regularize the model and prevent overfitting. We employed a discount factor of $\gamma=0.99$ to balance immediate and future rewards, and an estimation step of 3 for the agent to look ahead when estimating the expected return to improve the stability of Q-value estimates. The target network was updated every 300 steps and the buffer size was set to 20,000 to ensure sufficient experience storage for stable Q-value. A batch size of 128 was chosen to ensure efficient gradient updates and the model was trained for 200 epochs, each consisting of at least 10 episodes.

We consider two cases. The first case consists of uniform scenarios for all BS in the episode. This case can be further split into three subcases – all BSs are either exactbooked, underbooked, or overbooked. The second considers a more realistic situation where the scenario changes randomly for all BSs, yielding a randomly mixed distribution of scenarios (Mixed Booking). We demonstrate how the suggested approach reduces the total cost of reservations by validating it in all cases. The proposed model utilizes a DDQN model with 4 hidden layers, each having 128 channels, which is trained once offline on a server then gets deployed in vehicles. The computational requirements of the model are 101.38 KMAC and 202.75 KByte VRAM. All experiments were trained and tested on a Ubuntu server (CPU: AMD Ryzen 9 3950X 16-Core 3.5 GHz, RAM: 64 GB DDR4, GPU: NVIDIA RTX 2080 Ti). When compared to the NVIDIA AGX Xavier, which is recognized as the most powerful System-on-Chip (SoC)  \cite{Nvidia1}, \cite{liu2020computing}, the algorithm utilization is only 0.000004\% of the on-borad computation power of an autonomous vehicle, which can be seen as negligible. 
\subsection{Baselines}

We employed several baselines to compare the performance of our proposed approach. The first baseline involves using the original prereservation prices at $t_0$. We refer to this as the "No Policy" baseline because it does not employ any policy to update reservations, as seen in similar studies such as \cite{zang2019filling}, \cite{al2022optimal}, and \cite{zang2021soar}.
The second algorithm used is the "Greedy" algorithm, which updates reservations whenever there is a lower available price than the current one, as demonstrated in \cite{al2022heuristic} and \cite{chaisiri2011optimization}.
In addition to our proposed DDQN, we also trained and evaluated two other Q-learning methods: DQN and Dueling Deep Q-Learning Network (DuelingDQN).

\begin{figure}[t]
\centering
\includegraphics[width=0.46\textwidth]{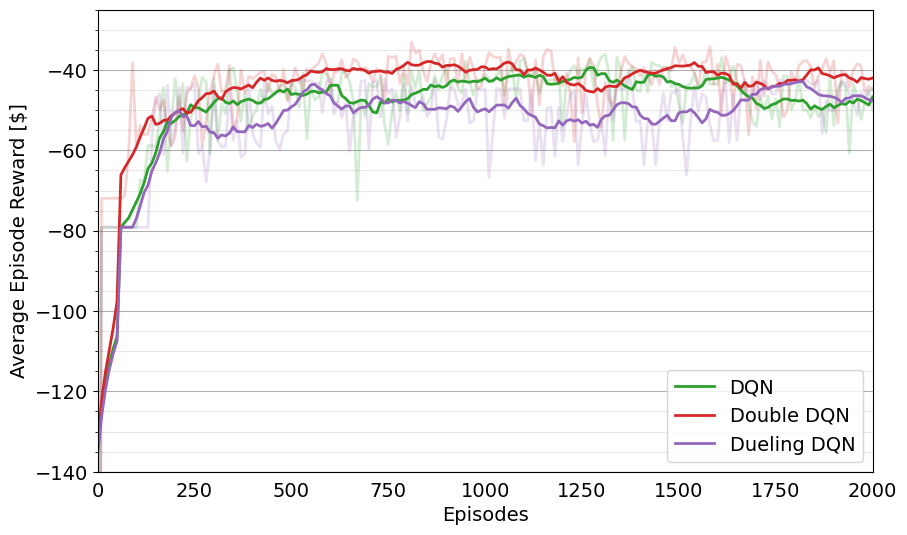}
\caption{Learning curves showing the convergence of the different DQN methods, demonstrating that the suggested Double DQN algorithm yields the highest average episode rewards.}
\label{fig:learning_curve}
\end{figure}

\subsection{Dataset} 
Dynamic pricing can be advantageous for both MNOs and vehicles. MNOs can optimize revenue and resources by adjusting prices based on demand, while vehicles can access resources at lower prices during low-demand periods. To evaluate our methodology, we employed a historical dataset of Amazon spot prices that fluctuate based on factors such as capacity, demand, geographic location, and specific instance type\cite{amazonspot}. For time-sensitive applications, vehicles require computing instances and communication links (i.e., bandwidth). As per our assumption, the prices to set up computing resources and communication links are same as that of spot pricing in Amazon, similar to the one referred in \cite{zang2021soar}.
In this study, we utilized prices from all instances and two regions (us-west-1b and us-west-1c) between April 17 and May 2 for training and May 3 to May 8 for testing purposes. To ensure the efficiency of the reservation system, we do not offer a free cancellation fee in our experiments. Instead, we have adopted a fee of 12\%, following the cancellation strategy fees used by Microsoft Azure \cite{Microsoft}.

\subsection{Experimental Results}
\begin{figure*}[t]
    \centering
    \subfloat[Cumulative costs for exactbooking]{\includegraphics[width=0.41\textwidth]{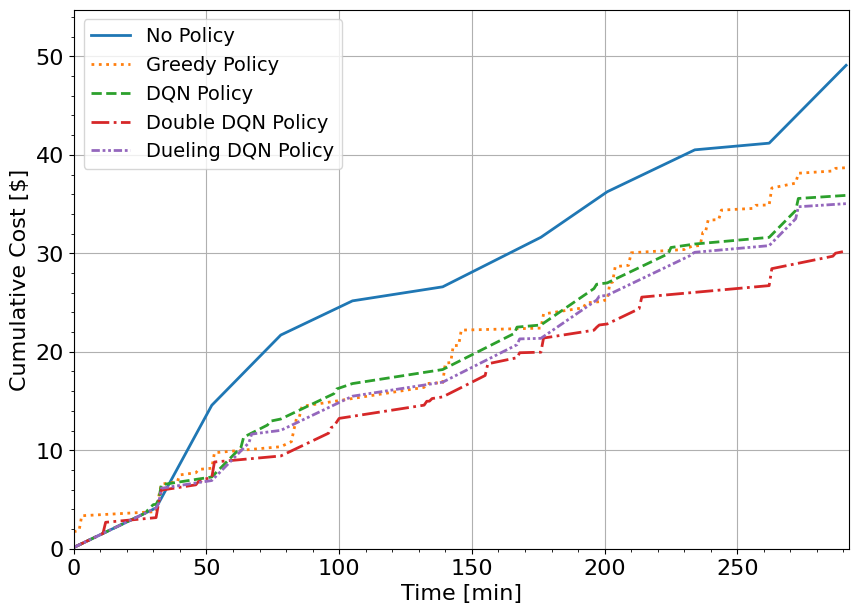}}
    \hspace{49pt}
    \subfloat[Cumulative costs for underbooking]{\includegraphics[width=0.41\textwidth]{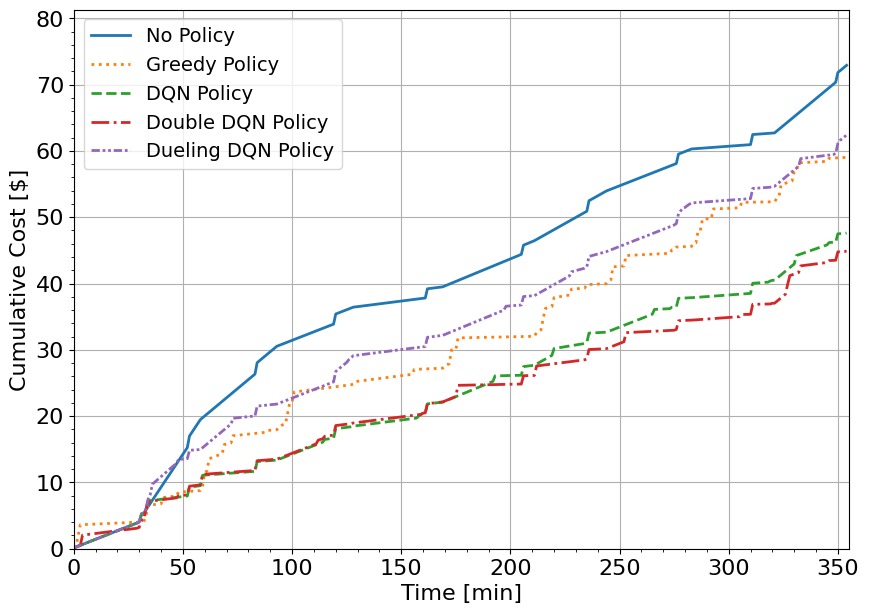}}%

    \centering
    \subfloat[Cumulative costs for overbooking]{\includegraphics[width=0.41\textwidth]{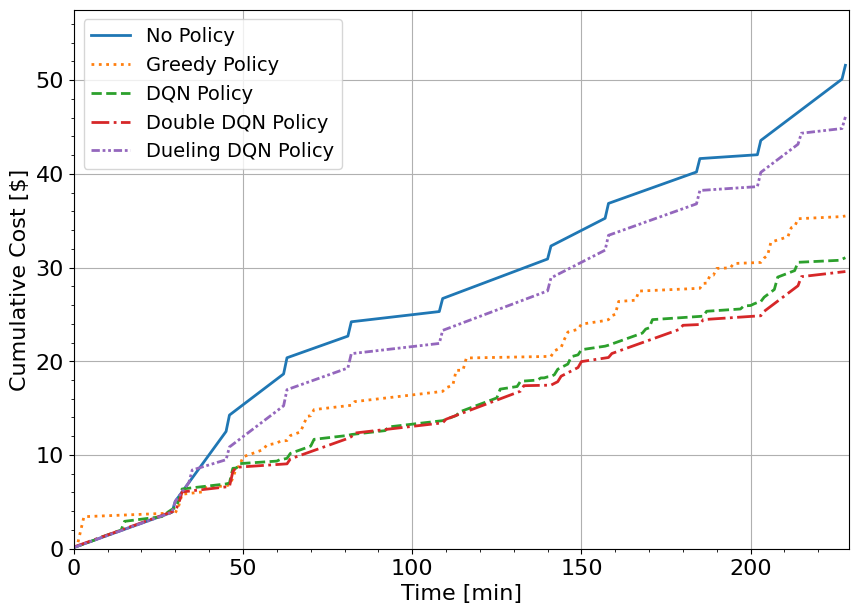}}%
    \hspace{49pt}
    \subfloat[Cumulative costs for mixedbooking]{\includegraphics[width=0.41\textwidth]{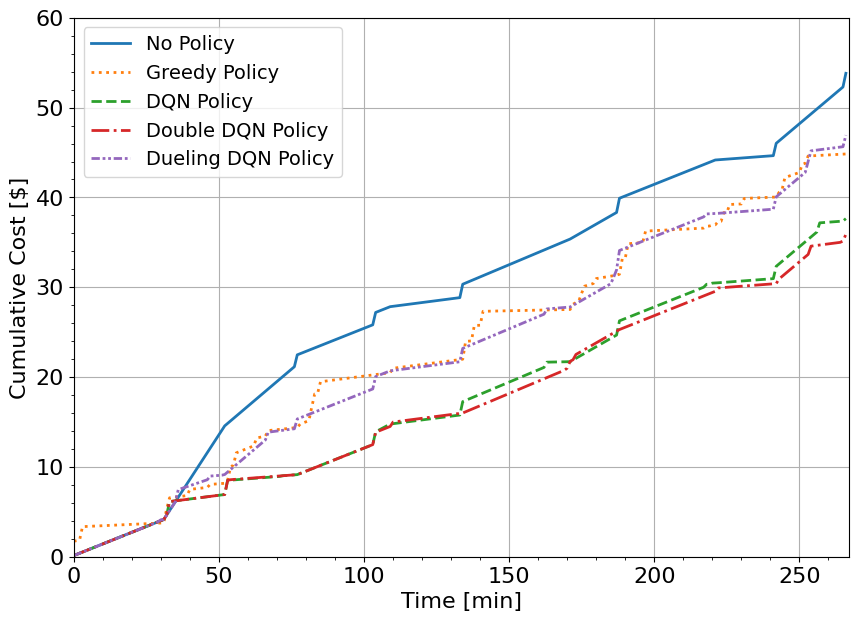}}%
    \caption{Cumulative episode rewards are significantly lower with suggested algorithms compared to baseline approaches across various scenarios: Exact booking (a), underbooking (b), overbooking (c), and mixed booking (d). The figure also illustrates the cumulative reservation costs for all scenarios.}
    \label{fig:episode_results}
\end{figure*}

The performance of the DRL algorithm and comparison of the outcomes with baseline algorithms have been presented in this section. Our objective is to illustrate how the algorithm reduces cumulative reservation costs and average costs incurred in situations involving variable booking. The convergence of the average cumulative reward during the trained episodes of the various DRL agents during training is depicted in Fig. \ref{fig:learning_curve}. The proposed agents were trained encountering various prices, base station numbers, and reservation lengths. The results, as demonstrated in the provided figure, indicate that the agents successfully converged on the optimal policy after 500 training episodes, learning the best approach to solve the bandwidth reservation problem.
Fig. \ref{fig:episode_results} a), b), c) and d) demonstrate that our suggested algorithms significantly outperform the baseline approaches in the cumulative reservation costs over time in all scenarios. The dynamic price and updates are not considered in the "No Policy" method, thus leading to high costs. Also, "Greedy Policy" updates every time there is a lower available price, entailing a higher cost caused by the cumulating cancellation fees. The proposed algorithm achieves better cumulative cost than both of these. Regarding the DQN methods, the plots clearly show the superiority of DDQL over DQN and Dueling DQN, which can be imputed to the decoupling of the action selection and value estimation to address the over-estimation issue.

\begin{figure}[t]
\centering
\includegraphics[width=0.42\textwidth]{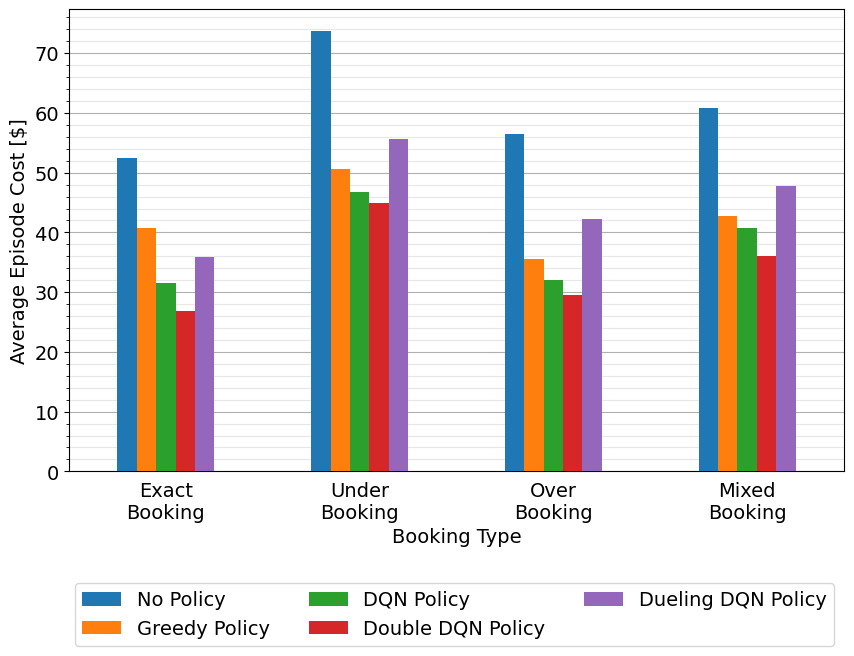}
\caption{Average costs per episode for all scenarios and policies. According to the figure, the suggested Double DQN policy has the lowest cost in all scenarios.}
\label{fig:avg_cost}
\end{figure}

In order to further evaluate the performance of the proposed algorithms, additional metrics were employed in the analysis. Specifically, the average episode cost for each scenario was calculated in order to identify the costs incurred by each algorithm in each scenario (Fig. \ref{fig:avg_cost}). This analysis provides further evidence for the superiority of the Double DQN method in comparison to other methods.

The results of this analysis demonstrate that the Double DQN policy is able to minimize the overall costs of the bandwidth reservation problem in all the scenarios. This suggests the Double DQN policy to be a highly effective method for solving the cost minimalization problem. The use of average episode cost as a metric allowed us to objectively compare the performance of different algorithms across different scenarios.

\begin{figure}[!t]
\centering
\begin{minipage}[t]{0.42\linewidth}
    \subfloat[]{\includegraphics[width=1\textwidth]{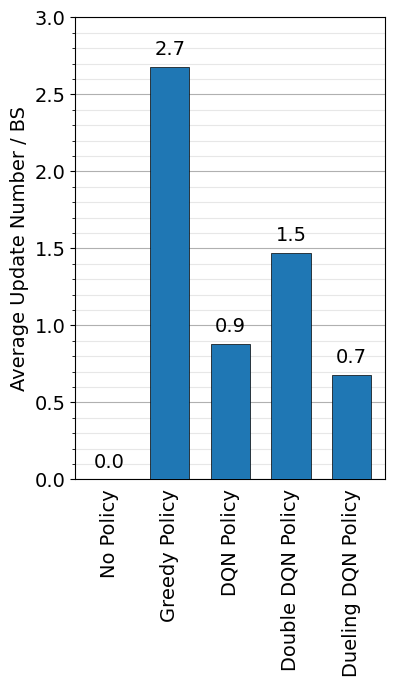}}
\end{minipage}%
    \hspace{3pt}
\begin{minipage}[t]{0.42\linewidth}
    \subfloat[]{\includegraphics[width=1\textwidth]{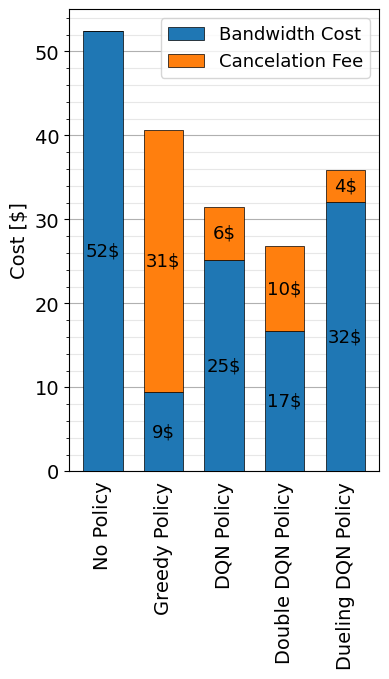}}
\end{minipage} 
\caption{The average update number of updates per BS (a) the cost components (cancelation and bandwidth) accounting for the paid full cost (b) for exactbooking. We can see that using a medium number of updates Double DQN has the best balance between the two cost components among the examined methods.}
\label{fig:double_figure}
\end{figure}

We also assessed the efficiency of the methods by analyzing the number of average updates and the ratio of bandwidth prices and cancellation costs per BS conducted for the exactbooking scenario (Fig. \ref{fig:double_figure}). Results showed that the greedy policy resulted in a high number of updates and high cancellation fees, leading to its higher overall cost, while DQN methods used fewer updates, yielding higher average prices but low cancelation fees. The Double DQN policy, among the DQN approaches, was found the best trade-off between bandwidth price and cancellation cost, resulting in the lowest overall cost. The analysis further supports the effectiveness of the Double DQN policy in optimizing bandwidth reservation.

\begin{figure}[t]
\centering
{\includegraphics[width=0.5\textwidth]{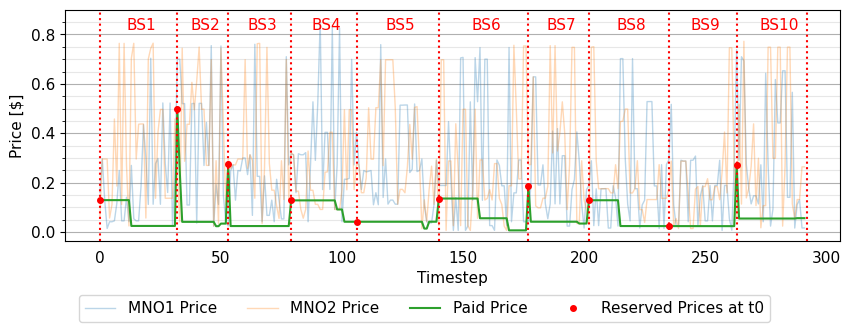}}

\caption{The reservation update decisions of the agent for an episode route consisting of 10 BSs with the available MNO prices. At each BS the agent starts with the initially reserved price and at each timestep it can decide to update it. Each jump in the paid price plot means an update decision.}
\label{fig:available_no_prices}
\end{figure}

Besides qualitative measurements of the agent's performance, we examined the actual MNO prices and the agent's actions to update to a lower price. The results, as shown in Fig. \ref{fig:available_no_prices} indicate that the algorithm generally performs well, quickly switching to lower prices at the onset of $BS_2$, $BS_3$, $BS_7$, and $BS_{10}$. However, there are instances where the agent decides to wait instead of instantly switching to a lower price, even if a previously reserved price at $t_0$ is higher. This decision is based on the consideration that the cancelation cost plus the bandwidth price is higher than the current price, hence the agent waits to switch later when it would be more advantageous. Even if this policy results in small periods of higher prices, it yields lower overall costs.

\begin{figure}[t]
\centering
{\includegraphics[width=0.5\textwidth]{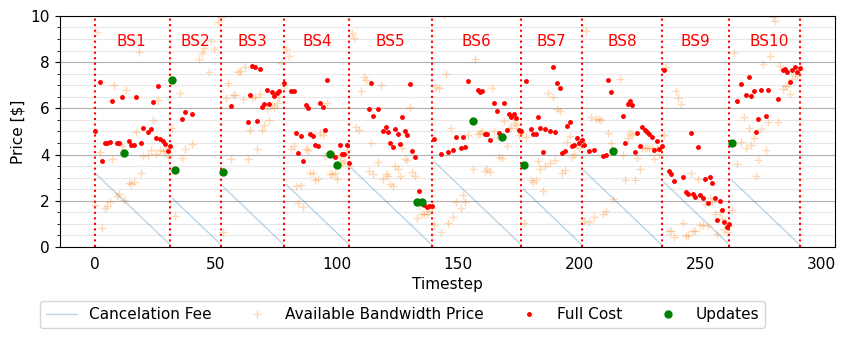}}
\caption{All reservation update possibilities and their yielded costs, comprising the cancelation fee and the currently available MNO price. Despite having access to only the current bandwidth prices, the agent consistently selects the option with the lowest available cost for the BS.}
\label{fig:price_paid}
\end{figure}

This is particularly evident in Fig. \ref{fig:price_paid}, where we can see the full prices that would have been paid for each BS in the event that the agent decided to switch at different times to different available prices (red points), and the influence of its components (cancelation fee: blue line, full BS bandwidth cost at alternative prices: yellow "+"). While the cumulative cancelation fee decreases with time, as there is less bandwidth time to cancel; the cumulative paid bandwidth fee increases, as staying on a more expensive MNO (at a previously reserved higher price) for a larger part of the BS as time is advancing. As the BS length, preserved price, and available prices change, the relationship between these two factors creates a complex function. 

\begin{figure*}[t]
\centering
\subfloat[Greedy policy]{\includegraphics[width=0.48\textwidth]
{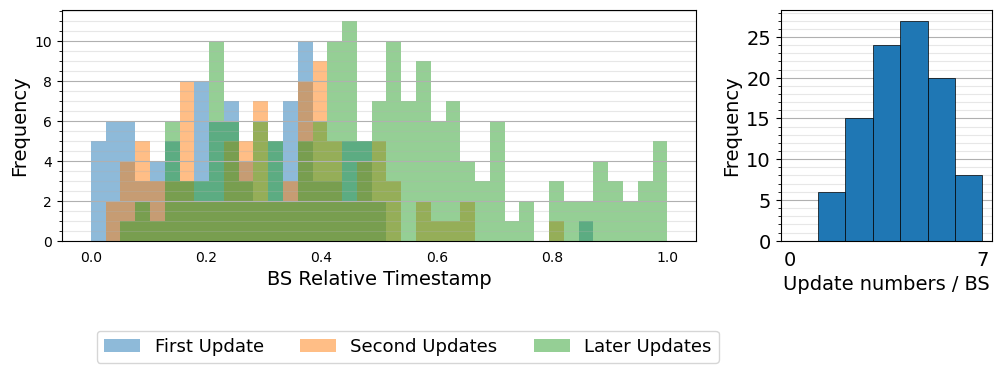}}
\hspace{1pt}
\subfloat[DQN policy]{\includegraphics[width=0.48\textwidth]{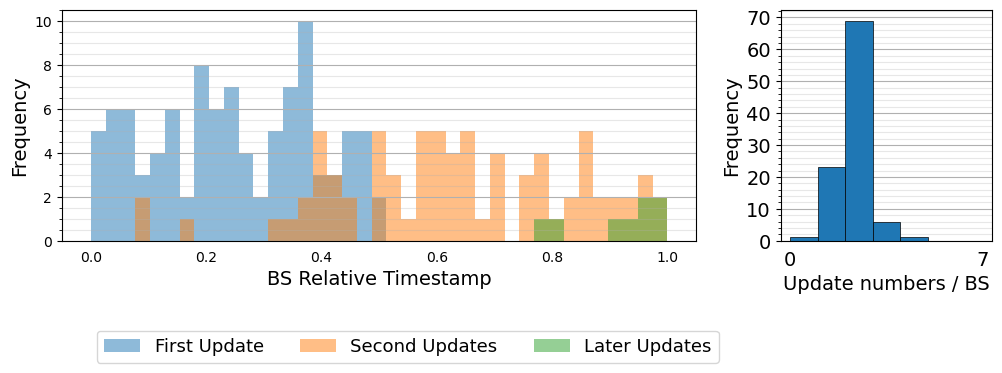}}
\hfill
\subfloat[Double DQN policy]{\includegraphics[width=0.48\textwidth]{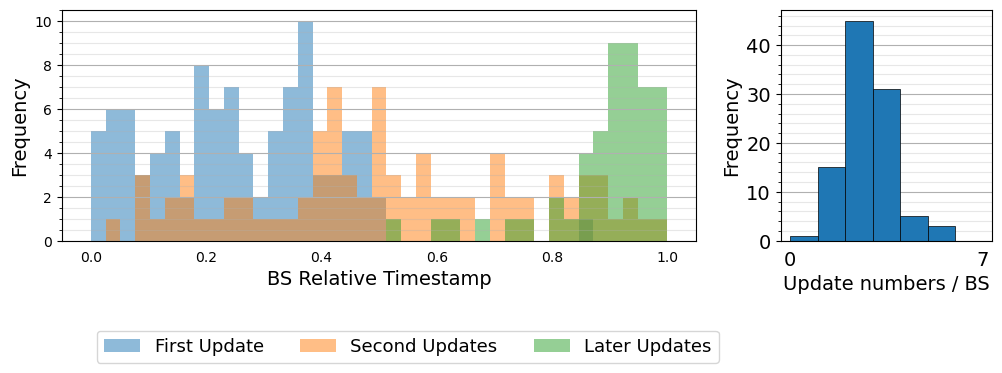}}
\hspace{1pt}
\subfloat[Dueling DQN policy]{\includegraphics[width=0.48\textwidth]{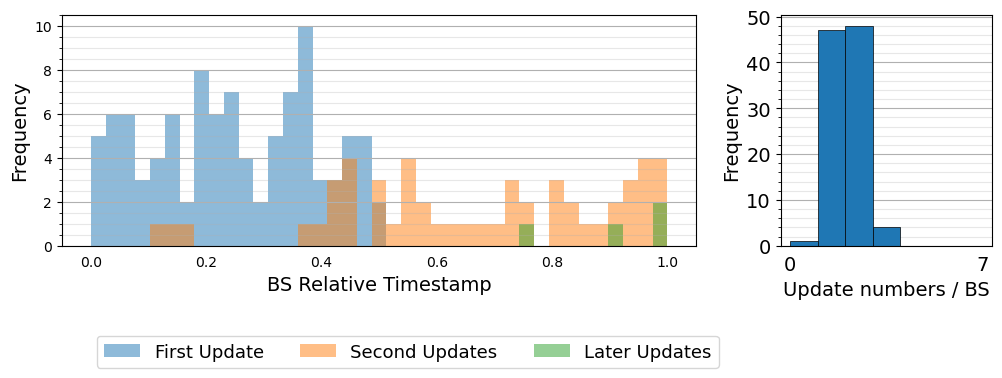}}
\caption{For all policies the left charts display the update time distribution relative to the time in the BS. Compared to the others, the Double DQNs make additional later updates towards the end of the BS when the cancellation fee is less to lower overall costs.
On the right charts, we can see the distribution of the number of updates in a BS. Compared to greedy and other DQN policies, the Double DQN policy makes a medium number of updates per BS.}
\label{fig:update_times_numbers}
\end{figure*}

Our analysis revealed that the agent, despite having access only to the currently available prices, consistently chooses the option corresponding to the minimal full cost for the BS. Additionally, if a lower price is available the agent often updates again close to the end of the BSs, where the cancelation fee is smaller. This can be seen in Fig. \ref{fig:update_times_numbers}: the first update is exclusively made in the first half of the BS, but a considerable amount of the later updates are made in the last 20\% of the BS. We can see on the right graph that this happens in most of the BSs: most of the times the agent makes one or even two correcting updates in a BS. Comparing it to other policies, they are less frequently following this policy. DQN and Dueling DQN also do secondary updates, but usually miss to do third updates, missing to lower their cost close to the end of the BSs (where cancelation updates are low) leading to their higher cost. On the other hand, the greedy policy makes a lot of later updates, whose cancelation costs lead to its higher overall cost.

We also observed that in some instances, it is more advantageous for the agent to wait for a later time to update to a lower price, but in other instances, especially if the prereserved price at $t_0$ is high, the agent switches to a lower price as soon as it enters the BS, resulting in large spikes at the start of some BSs.

Based on these findings, we propose that allowing the agent to cancel reservations and update to lower prices before entering new BSs may help to mitigate the spikes in prices and minimize overall costs. This serves as a potential area for future research and improvement of the algorithm.
\section{CONCLUSION}
In this article, we studied the potential cost optimization for vehicular bandwidth reservation in dynamically priced multi-operator networks. We proposed a multi-objective reinforcement learning-based strategy to update bandwidth reservations upon relevant price changes and address the challenges of underbooking and overbooking in a cost-effective way. The problem was formalized as a Markov Decision Model used to train a Double Deep Q-learning reinforcement learning model on a real-world dataset. The results show that our proposed model outperforms existing strategies. Our model reduces the bandwidth costs by 40\% across all scenarios when compared to a control group lacking a policy and effectively solves underbooking and overbooking scenarios at the lowest cost.

This study demonstrates the effectiveness of using a reinforcement learning-based strategy to address the challenges of bandwidth cost minimization and solving underbooking and overbooking, providing a promising cost-effective solution for the multi-operator vehicular bandwidth reservation problem. In future work, we aim to focus on more elaborate environments with dynamic BS ranges and varying MNO numbers to enable the development of more realistic and intelligent reservation update strategies, leading to enhanced resource allocation and cost minimization in dynamic vehicular networking environments.

\bibliography{Main2}

\begin{thebibliography}{10}
\providecommand{\url}[1]{#1}
\csname url@samestyle\endcsname
\providecommand{\newblock}{\relax}
\providecommand{\bibinfo}[2]{#2}
\providecommand{\BIBentrySTDinterwordspacing}{\spaceskip=0pt\relax}
\providecommand{\BIBentryALTinterwordstretchfactor}{4}
\providecommand{\BIBentryALTinterwordspacing}{\spaceskip=\fontdimen2\font plus
\BIBentryALTinterwordstretchfactor\fontdimen3\font minus \fontdimen4\font\relax}
\providecommand{\BIBforeignlanguage}[2]{{%
\expandafter\ifx\csname l@#1\endcsname\relax
\typeout{** WARNING: IEEEtran.bst: No hyphenation pattern has been}%
\typeout{** loaded for the language `#1'. Using the pattern for}%
\typeout{** the default language instead.}%
\else
\language=\csname l@#1\endcsname
\fi
#2}}
\providecommand{\BIBdecl}{\relax}
\BIBdecl

\bibitem{Nvidia}
\BIBentryALTinterwordspacing
Nvidia drive thor. [Online]. Available: \url{https://www.nvidia.com/en-us/self-driving-cars/hardware/}
\BIBentrySTDinterwordspacing

\bibitem{chen2021energy}
X.~Chen and G.~Liu, ``Energy-efficient task offloading and resource allocation via deep reinforcement learning for augmented reality in mobile edge networks,'' \emph{IEEE Internet Things J.}, vol.~8, no.~13, pp. 10\,843--10\,856, Jul. 2021.

\bibitem{cheng2021multiagent}
Z.~Cheng, M.~Min, M.~Liwang, L.~Huang, and Z.~Gao, ``Multiagent ddpg-based joint task partitioning and power control in fog computing networks,'' \emph{IEEE Internet Things J.}, vol.~9, no.~1, pp. 104--116, Jan. 2022.

\bibitem{chiang2016fog}
M.~Chiang and T.~Zhang, ``Fog and iot: An overview of research opportunities,'' \emph{IEEE Internet Things J.}, vol.~3, no.~6, pp. 854--864, Dec. 2016.

\bibitem{al2024resources}
A.~Al-Khatib, S.~Ehsanfar, K.~Moessner, and H.~Timinger, ``Resources reservation schemes for time-sensitive networked vehicular applications with a view on isac,'' \emph{IEEE Access}, 2024.

\bibitem{meneguette2021vehicular}
R.~Meneguette, R.~De~Grande, J.~Ueyama, G.~P.~R. Filho, and E.~Madeira, ``Vehicular edge computing: Architecture, resource management, security, and challenges,'' \emph{ACM Comput. Surv. (CSUR)}, vol.~55, no.~1, pp. 1--46, Nov. 2021.

\bibitem{al2020priority}
A.~Al-Khatib and A.~Khelil, ``Priority- and reservation-based slicing for future vehicular networks,'' in \emph{Proc. IEEE Conf. Netw. Softwa. (NetSoft)}, Aug. 2020, pp. 36--42.

\bibitem{al2021bandwidth}
A.~Al-Khatib, A.~Khelil, and M.~Balfaqih, ``Bandwidth slicing with reservation capability and application priority awareness for future vehicular networks,'' in \emph{Proc. Advanced Info. Netw. Applications Conf. (AINA)}, Apr. 2021, pp. 681--691.

\bibitem{niyato2008competitive}
D.~Niyato and E.~Hossain, ``Competitive spectrum sharing in cognitive radio networks: a dynamic game approach,'' \emph{IEEE Trans. Wireless Commun.}, vol.~7, no.~7, pp. 2651--2660, Jul. 2008.

\bibitem{chen2020edge}
D.~Chen, Y.-C. Liu, B.~Kim, J.~Xie, C.~S. Hong, and Z.~Han, ``Edge computing resources reservation in vehicular networks: A meta-learning approach,'' \emph{IEEE Trans. Veh. Technol.}, vol.~69, no.~5, pp. 5634--5646, May. 2020.

\bibitem{zang2019filling}
S.~Zang, W.~Bao, P.~L. Yeoh, B.~Vucetic, and Y.~Li, ``Filling two needs with one deed: Combo pricing plans for computing-intensive multimedia applications,'' \emph{IEEE J. Sel. Areas Commun.}, vol.~37, no.~7, pp. 1518--1533, May. 2019.

\bibitem{al2022optimal}
A.~Al-Khatib, F.~Al-Khateeb, A.~Khelil, and K.~Moessner, ``Optimal timing for bandwidth reservation for time-sensitive vehicular applications,'' in \emph{Proc. IEEE Int. Conf. Fog Edge Comput. (ICFEC)}, May. 2022, pp. 94--99.

\bibitem{zang2021soar}
S.~Zang, W.~Bao, P.~L. Yeoh, B.~Vucetic, and Y.~Li, ``Soar: Smart online aggregated reservation for mobile edge computing brokerage services,'' \emph{IEEE Trans. Mobile Comput.}, vol.~22, no.~1, pp. 527--540, Jan. 2023.

\bibitem{al2024optimizing}
A.~Al-Khatib, K.~Moessner, and H.~Timinger, ``Optimizing bandwidth reservation decision time in vehicular networks using batched lstm.'' \emph{Int. J. Adv. Comput. Sci. Appl.}, vol.~15, no.~2, 2024.

\bibitem{al2024blockchain}
A.~Al-Khatib, H.~Hadi, H.~Timinger, and K.~Moessner, ``Blockchain-empowered resource trading for optimizing bandwidth reservation in vehicular networks,'' \emph{IEEE Access}, 2024.

\bibitem{AmazonW}
\BIBentryALTinterwordspacing
Aws pricing - how does aws pricing work? [Online]. Available: \url{https://aws.amazon.com/pricing/?nc1=h ls}
\BIBentrySTDinterwordspacing

\bibitem{Atandt}
\BIBentryALTinterwordspacing
At\&t, wireless plans. [Online]. Available: \url{https://www.att.com/plans/wireless/}
\BIBentrySTDinterwordspacing

\bibitem{han2018dynamic}
D.~Han, W.~Chen, and Y.~Fang, ``A dynamic pricing strategy for vehicle assisted mobile edge computing systems,'' \emph{IEEE Wireless Commun. Lett.}, vol.~8, no.~2, pp. 420--423, Apr. 2018.

\bibitem{liao2021intelligent}
Y.~Liao, X.~Qiao, Q.~Yu, and Q.~Liu, ``Intelligent dynamic service pricing strategy for multi-user vehicle-aided mec networks,'' \emph{Future Generation Comput. Sys.}, vol. 114, pp. 15--22, Jan. 2021.

\bibitem{luong2016data}
N.~C. Luong, D.~T. Hoang, P.~Wang, D.~Niyato, D.~I. Kim, and Z.~Han, ``Data collection and wireless communication in internet of things (iot) using economic analysis and pricing models: A survey,'' \emph{IEEE Commun. Surv. Tuts.}, vol.~18, no.~4, pp. 2546--2590, Jun. 2016.

\bibitem{amazonspot}
\BIBentryALTinterwordspacing
Amazon ec2 spot instances pricing. [Online]. Available: \url{https://aws.amazon.com/ec2/spot/pricing/}
\BIBentrySTDinterwordspacing

\bibitem{lin2020backup}
L.~Lin, L.~Pan, and S.~Liu, ``Backup or not: An online cost optimal algorithm for data analysis jobs using spot instances,'' \emph{IEEE Access}, vol.~8, pp. 144\,945--144\,956, Aug. 2020.

\bibitem{sen2013survey}
S.~Sen, C.~Joe-Wong, S.~Ha, and M.~Chiang, ``A survey of smart data pricing: Past proposals, current plans, and future trends,'' \emph{ACM Comput. Surveys}, vol.~46, no.~2, pp. 1--37, Nov. 2013.

\bibitem{al2022heuristic}
A.~Al-Khatib, M.~U. Hassan, and K.~Moessner, ``Heuristic optimization of bandwidth reservation cost for vehicular applications,'' in \emph{Proc. IEEE Global Commun. Conf. (GLOBECOM)}, Dec. 2022, pp. 4909--4915.

\bibitem{chaisiri2011optimization}
S.~Chaisiri, B.-S. Lee, and D.~Niyato, ``Optimization of resource provisioning cost in cloud computing,'' \emph{IEEE Trans. Services Comput.}, vol.~5, no.~2, pp. 164--177, Feb. 2011.

\bibitem{olariu2019survey}
S.~Olariu, ``A survey of vehicular cloud research: Trends, applications and challenges,'' \emph{IEEE Trans. Intell. Transp. Syst.}, vol.~21, no.~6, pp. 2648--2663, Jun. 2020.

\bibitem{meneguette2018avarac}
R.~I. Meneguette, A.~Boukerche, and A.~H. Pimenta, ``Avarac: An availability-based resource allocation scheme for vehicular cloud,'' \emph{IEEE Trans. Intell. Transp. Syst.}, vol.~20, no.~10, pp. 3688--3699, Oct. 2019.

\bibitem{ghazizadeh2015reasoning}
P.~Ghazizadeh, R.~Florin, A.~G. Zadeh, and S.~Olariu, ``Reasoning about mean time to failure in vehicular clouds,'' \emph{IEEE Trans. Intell. Transp. Syst.}, vol.~17, no.~3, pp. 751--761, Mar. 2016.

\bibitem{zhang2023dynamic}
J.~Zhang, S.~Chen, X.~Wang, and Y.~Zhu, ``Dynamic reservation of edge servers via deep reinforcement learning for connected vehicles,'' \emph{IEEE Trans. Mobile Comput.}, vol.~22, no.~5, pp. 2661--2674, May. 2023.

\bibitem{chen2018optimized}
X.~Chen, H.~Zhang, C.~Wu, S.~Mao, Y.~Ji, and M.~Bennis, ``Optimized computation offloading performance in virtual edge computing systems via deep reinforcement learning,'' \emph{IEEE Internet Things J.}, vol.~6, no.~3, pp. 4005--4018, Jun. 2019.

\bibitem{he2017deep}
Y.~He, Z.~Zhang, F.~R. Yu, N.~Zhao, H.~Yin, V.~C. Leung, and Y.~Zhang, ``Deep-reinforcement-learning-based optimization for cache-enabled opportunistic interference alignment wireless networks,'' \emph{IEEE Trans. Veh. Technol.}, vol.~66, no.~11, pp. 10\,433--10\,445, Nov. 2017.

\bibitem{van2016deep}
H.~Van~Hasselt, A.~Guez, and D.~Silver, ``Deep reinforcement learning with double q-learning,'' in \emph{Proc. Conf. Artifi. Intell. (AAAI)}, vol.~30, no.~1, Mar. 2016.

\bibitem{yu2012resource}
X.~Yu, P.~Navaratnam, and K.~Moessner, ``Resource reservation schemes for ieee 802.11-based wireless networks: A survey,'' \emph{IEEE Commun. Surv. Tuts.}, vol.~15, no.~3, pp. 1042--1061, Nov. 2012.

\bibitem{cao2016share}
Y.~Cao, C.~Long, T.~Jiang, and S.~Mao, ``Share communication and computation resources on mobile devices: A social awareness perspective,'' \emph{IEEE Wireless Commun.}, vol.~23, no.~4, pp. 52--59, Aug. 2016.

\bibitem{chen2020stackelberg}
Y.~Chen, Z.~Li, B.~Yang, K.~Nai, and K.~Li, ``A stackelberg game approach to multiple resources allocation and pricing in mobile edge computing,'' \emph{Future Generation Comput. Sys.}, vol. 108, pp. 273--287, Jul. 2020.

\bibitem{bajaj2015spectrum}
I.~Bajaj, Y.~H. Lee, and Y.~Gong, ``A spectrum trading scheme for licensed user incentives,'' \emph{IEEE Trans. Commun.}, vol.~63, no.~11, pp. 4026--4036, Nov. 2015.

\bibitem{al2025bandwidth}
A.~Al-Khatib, A.~Ahmed, K.~Moessner, and H.~Timinger, ``Bandwidth reservation for time-critical vehicular applications: A multi-operator environment,'' \emph{IEEE Trans. Consum. Electron.}, May 2025.

\bibitem{nadembega2012destination}
A.~Nadembega, T.~Taleb, and A.~Hafid, ``A destination prediction model based on historical data, contextual knowledge and spatial conceptual maps,'' in \emph{Proc. IEEE Int. Conf. Commun. (ICC)}, Jun. 2012, pp. 1416--1420.

\bibitem{nadembega2014mobility}
A.~Nadembega, A.~Hafid, and T.~Taleb, ``Mobility-prediction-aware bandwidth reservation scheme for mobile networks,'' \emph{IEEE Trans. Veh. Technol.}, vol.~64, no.~6, pp. 2561--2576, Jun. 2015.

\bibitem{soh2006predictive}
W.-S. Soh and H.~S. Kim, ``A predictive bandwidth reservation scheme using mobile positioning and road topology information,'' \emph{IEEE/ACM Trans. Netw.}, vol.~14, pp. 1078--1091, Oct. 2006.

\bibitem{choi2002adaptive}
S.~Choi and K.~G. Shin, ``Adaptive bandwidth reservation and admission control in qos-sensitive cellular networks,'' \emph{IEEE Trans. Parallel and Distributed Syst.}, vol.~13, pp. 882--897, Sep. 2002.

\bibitem{liwang2022overbooking}
M.~Liwang and X.~Wang, ``Overbooking-empowered computing resource provisioning in cloud-aided mobile edge networks,'' \emph{IEEE/ACM Trans. Netw.}, vol.~30, no.~5, pp. 2289--2303, Oct. 2022.

\bibitem{Microsoft}
\BIBentryALTinterwordspacing
Azure reserved virtual machine instances. [Online]. Available: \url{https://azure.microsoft.com/en- us/pricing/reserved-vm-instances/}
\BIBentrySTDinterwordspacing

\bibitem{sutton2018reinforcement}
R.~S. Sutton and A.~G. Barto, \emph{Reinforcement learning: An introduction}.\hskip 1em plus 0.5em minus 0.4em\relax MIT Press, 2018.

\bibitem{zhang2020deep}
X.~Zhang, Y.~Xiao, Q.~Li, and W.~Saad, ``Deep reinforcement learning for fog computing-based vehicular system with multi-operator support,'' in \emph{Proc. IEEE Int. Conf. Commun. (ICC)}, Jun. 2020, pp. 1--6.

\bibitem{watkins1989learning}
C.~J. Watkins, ``Learning from delayed rewards,'' Ph.D. dissertation, Royal Holloway, University of London, 1989.

\bibitem{Nvidia1}
\BIBentryALTinterwordspacing
Nvidia jetson agx xavier delivers 32 teraops for new era of ai in robotics. [Online]. Available: \url{https://developer.nvidia.com/blog/nvidia-jetson-agx-xavier-32-teraops-ai-robotics/}
\BIBentrySTDinterwordspacing

\bibitem{liu2020computing}
L.~Liu, S.~Lu, R.~Zhong, B.~Wu, Y.~Yao, Q.~Zhang, and W.~Shi, ``Computing systems for autonomous driving: State of the art and challenges,'' \emph{IEEE Internet Things J.}, vol.~8, no.~8, pp. 6469--6486, Apr. 2020.

\end{thebibliography}

\end{document}